\documentclass[aps,prl,twocolumn,showpacs,superscriptaddress]{revtex4-1} 

\usepackage{amsmath,amssymb,bm,wrapfig,mathrsfs}
\usepackage[dvipdfm,hyperfootnotes=false]{hyperref}
\usepackage[dvipdfmx]{graphicx,color}
\usepackage{dcolumn}   
\usepackage{amssymb}   
\usepackage{here}

\begin{document}

\title{Large Unidirectional Magnetoresistance in a Magnetic Topological Insulator}
\author {K. Yasuda}
\email{yasuda@cmr.t.u-tokyo.ac.jp}
\affiliation{Department of Applied Physics and Quantum-Phase Electronics Center (QPEC), University of Tokyo, Tokyo 113-8656, Japan}
\author {A. Tsukazaki}
\affiliation{Institute for Materials Research, Tohoku University, Sendai 980-8577, Japan}
\author {R. Yoshimi}
\affiliation{RIKEN Center for Emergent Matter Science (CEMS), Wako 351-0198, Japan}
\author {K. S. Takahashi}
\affiliation{RIKEN Center for Emergent Matter Science (CEMS), Wako 351-0198, Japan}
\author {M. Kawasaki}
\affiliation{Department of Applied Physics and Quantum-Phase Electronics Center (QPEC), University of Tokyo, Tokyo 113-8656, Japan}
\affiliation{RIKEN Center for Emergent Matter Science (CEMS), Wako 351-0198, Japan}
\author {Y. Tokura}
\affiliation{Department of Applied Physics and Quantum-Phase Electronics Center (QPEC), University of Tokyo, Tokyo 113-8656, Japan}
\affiliation{RIKEN Center for Emergent Matter Science (CEMS), Wako 351-0198, Japan}
\date{\today}

\begin{abstract}
We report current-direction dependent or unidirectional magnetoresistance (UMR) in magnetic/nonmagnetic topological insulator (TI) heterostructures, Cr$_x$(Bi$_{1-y}$Sb$_y$)$_{2-x}$Te$_3$/(Bi$_{1-y}$Sb$_y$)$_2$Te$_3$, that is several orders of magnitude larger than in other reported systems. From the magnetic field and temperature dependence, the UMR is identified to originate from the asymmetric scattering of electrons by magnons. In particular, the large magnitude of UMR is an outcome of spin-momentum locking and a small Fermi wavenumber at the surface of TI. In fact, the UMR is maximized around the Dirac point with the minimal Fermi wavenumber.
\end{abstract}

\maketitle

Transfer and conservation of angular momentum is at the heart of spintronics \cite{4,5,6,7,8}; spin transfer torque works to convert spin-polarized current to and from magnons, thus enabling the electrical control of magnetism. For example, current induced magnetization reversal and spin torque ferromagnetic resonance through spin injection have been realized in various heterostructures based on heavy metal element (with large spin-orbit coupling) and ferromagnet, such as Pt/Py, Ta/CoFeB and Pt/Co \cite{5,6,7,8}. Recently, it has been reported that unidirectional magnetoresistance (UMR) emerges in such heterostructures under in-plane magnetization \cite{1,2,3}; the resistance value is different depending on the sign of the outer product of current $J$ and magnetization $M$ vectors. There, the spin accumulation direction, either parallel or anti-parallel with $M$, at the interface by spin Hall effect has been proposed to be a major origin of UMR \cite{1,2,3}, in analogy to the giant magnetoresistance (GMR) effect \cite{9,10}, which depends on the relationship of magnetizations, parallel or antiparallel, in stacked ferromagnetic metal layers. In a broader context, the UMR is expected to be further enhanced in topological insulator (TI) with large spin Hall angle \cite{11,12,13,14} via spin-momentum locking (Fig. 1(a)), because the spin polarization at the surface would govern the amplitude of this effect.

In this study, we investigate the UMR of TI heterostructures \cite{12,15,16} composed of nonmagnetic TI (Bi$_{1-y}$Sb$_y$)$_2$Te$_3$ (BST) \cite{17,18} and magnetic TI Cr$_x$(Bi$_{1-y}$Sb$_y$)$_{2-x}$Te$_3$ (CBST) \cite{19} on insulating InP substrate. By tuning the composition $y$, we could control the Fermi energy $E_\mathrm{F}$ of the surface state inside the bulk band gap, which is confirmed by the Hall effect measurement \cite{15,16,17,18}. Here, the main players of conduction are top and bottom surfaces with single Dirac cones around the $\Gamma$ point \cite{15,16,17',17''}. In addition, in the heterostructure, only one surface involved in the Cr-doped layer effectively interacts with magnetism \cite{15,16}. Thus, in terms of the symmetry consideration of spin-momentum locking (Fig. 1(a)), we can expect that magnetoresistance depends on the relative configuration between surface electron spin and $M$ directions; parallel (Fig. 1(b)) or antiparallel (Fig. 1(c)).

Thin films of TI heterostructures were grown with molecular beam epitaxy in the same procedures as described in Refs. \cite{15} and \cite{16}. The nominal compositions of TI heterostructure Cr$_x$(Bi$_{1-y}$Sb$_y$)$_{2-x}$Te$_3$/(Bi$_{1-y}$Sb$_y$)$_2$Te$_3$ are $x \sim 0.2$ and $y \sim 0.86$. Using photolithography and Ar ion milling, thin films were patterned into the shape of Hall bar, 10 $\mu$m in width and 36 $\mu$m in length. After that, the electrodes Au (45 nm)/Ti (5 nm) were formed by electron beam deposition \cite{SI}. The transport measurements were performed mainly at 2 K in Physical Property Measurement System (Quantum Design) using dc current source and a voltmeter.

\begin{figure*}
\centering
\includegraphics{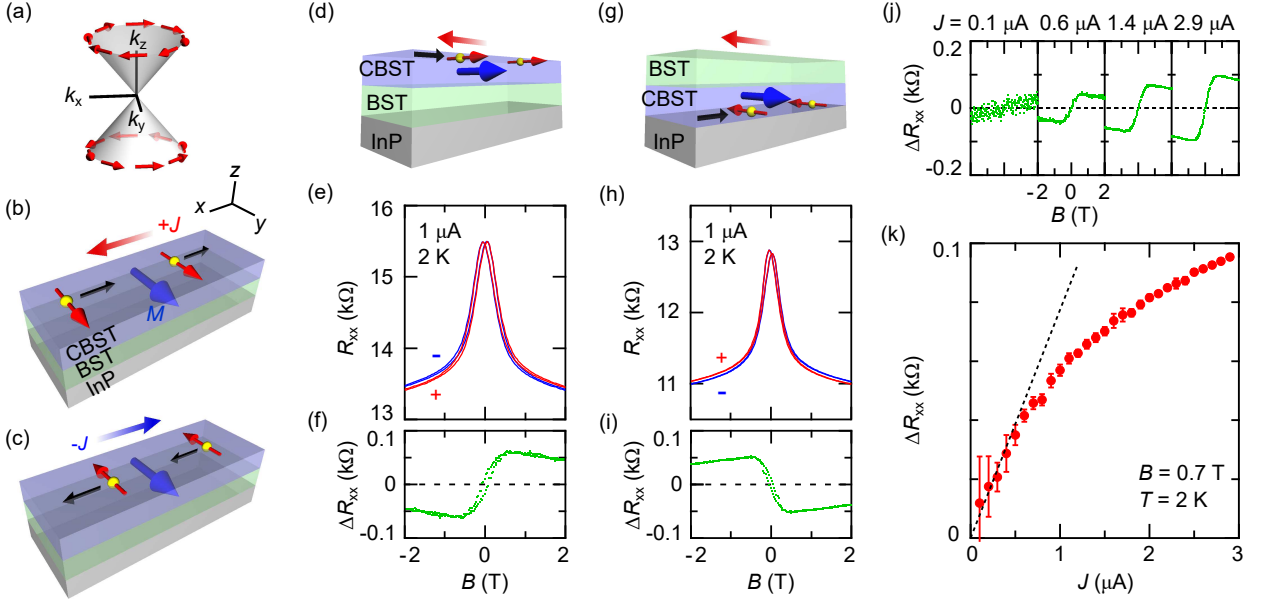}
\caption{(color online) (a) Schematic diagram of spin-momentum locking of surface Dirac state in TI. (b), (c) Schematic illustration of the concept for UMR in TI heterostructures Cr$_x$(Bi$_{1-y}$Sb$_y$)$_{2-x}$Te$_3$/(Bi$_{1-y}$Sb$_y$)$_2$Te$_3$ (CBST/BST) on InP substrate under $+J$ (b) and $-J$ (c) dc current. Here, magnetic field, magnetization and dc current are along the in-plane direction, where dc current is applied perpendicular to the magnetization direction. (d) Schematic illustration of a ``normal" CBST/BST heterostructure. (e) Magnetic field dependence of resistance $R_\mathrm{xx}$ for the sample depicted in (d) measured under $J = +1$ $\mu$A (red) and $J = -1$ $\mu$A (blue) at 2 K. (f), Difference of the resistance $\Delta R_\mathrm{xx}$ of plus and minus current shown in (e). (g)-(i) The same as (d)-(f) for the ``inverted" BST/CBST heterostructure. (j) $\Delta R_\mathrm{xx}$ measured under various current for the normal CBST/BST heterostructure. (k) Current $J$ dependence of $\Delta R_\mathrm{xx}$ at 2 K under $B = 0.7$ T for the normal CBST/BST. The black dotted line shows a slope in the low-$J$ region. }
\label{F1}
\end{figure*}

Figure 1(e) shows the measured magnetoresistance of the heterostructure CBST (3 nm)/BST (5 nm) (Fig. 1(d)). First, we notice that resistance decreases with increasing in-plane magnetic field $B$. Because of the out-of-plane anisotropy of $M$ in CBST, $M$ initially points along the $z$-direction forming the exchange gap in surface Dirac state. As magnetic field is applied up to 0.7 T, the magnetization direction gradually changes to the in-plane so that the eventual gap closing of the Dirac surface state causes negative magnetoresistance \cite{SI}. Also, we note that the resistance measured under +1 $\mu$A (red) and $-1$ $\mu$A (blue) at 2 K show a noticeable deviation as shown in Fig. 1(e); the difference $\Delta R_\mathrm{xx}$ between the two current directions is plotted in Fig. 1(f). Here, $\Delta R_\mathrm{xx}$ is anti-symmetrized as a function of $B$ and $M$. $\Delta R_\mathrm{xx}$ is initially almost zero at 0 T where $M$ is pointing out-of-plane, and then increases as the field increases up to 0.7 T. At higher magnetic field above 0.7 T, $\Delta R_\mathrm{xx}$ becomes almost constant, whose sign is reversed in accordance with $M$ reversal in CBST. Furthermore, $\Delta R_\mathrm{xx}$ is also reversed in sign, as shown in Figs. 1(h) and 1(i), for the inverted heterostructure BST (3 nm)/CBST (5 nm) (Fig. 1(g)), while showing the similar absolute magnitude of UMR. This is most likely because the manner of the spin-momentum locking is opposite between the top and bottom surfaces as depicted in Figs. 1(d) and 1(g). This leads to the cancellation of UMR in the case of the single-layer CBST film \cite{SI}. Figures 1(j) and 1(k) show the current amplitude dependence of UMR. While $\Delta R_\mathrm{xx}$ shows a negligibly small difference with current amplitude of 0.1 $\mu$A, it is progressively enhanced with increasing current. The current $J$ dependence of $\Delta R_\mathrm{xx}$ at 0.7 T is summarized in Fig. 1(k), which shows a linear relationship in a low current region, typically $J$ $<$ 0.5 $\mu$A. Therefore, the relationship between electric field $E_\mathrm{x}$ and current density $j_\mathrm{x}$ should be expressed in a nonlinear form in such a low current region, 
\begin{equation}
E_\mathrm{x}=R_\mathrm{xx} j_\mathrm{x}+R_\mathrm{xx}^{(2)} j_\mathrm{x}^2.
\end{equation}
Here, $\Delta R_\mathrm{xx}=2R_\mathrm{xx}^{(2)} j_\mathrm{x}$ is linearly proportional to current density. The derivation from the linear relationship in Fig. 1(k) at high current ($>$ 0.5 $\mu$A) is attributed to heating effect by fairly large current excitation, up to $\Delta T$ = 2.3 K at $J = 3\ \mu$A as estimated from the change of $R_\mathrm{xx}$ \cite{SI}. Hereafter, we applied $\pm1\ \mu$A for the measurements to get enough S/N ratio but to make the heating effect as small as possible.

In Table \ref{T1}, we compare the magnitude of UMR in the present device with those of previously reported heterostructures \cite{1,2,3}. Since it is linear in current, we adopt the quantity ($\Delta R_\mathrm{xx}$/$R_\mathrm{xx})/j$ as a measure of UMR for a fair comparison. In the TI heterostructure, we define the current density by considering each surface conduction thickness of $\sim$ 1 nm \cite{20} (see also the legend of Table 1). Even though the current density is much smaller than other systems, $\Delta R_\mathrm{xx}$/$R_\mathrm{xx}$ is comparable or larger. Therefore, the amplitude of ($\Delta R_\mathrm{xx}$/$R_\mathrm{xx})/j$ is quite large, $10^2-10^6$ times larger than other bilayer systems, $e.g.$ GaMnAs heterstructure or Pt/Co \cite{1,2,3}. 

\begin{table*}
\caption{\label{T1} Comparison of UMR magnitude for various heterostructures. Note that the values marked with asterisk $^*$ for CBST/BST are calculated with assuming the thickness of conductive region $\sim$ 2 nm of top and bottom surface states \cite{20}. Even if the total thickness of the whole film ($\sim$ 8 nm) were taken, the values would be changed only by a factor of four.\\}
\begin{tabular}{l|l l l l l}
Material & $j$ (A/cm$^2$)&  $R_\mathrm{xx}$ ($\Omega$) & $\Delta R_\mathrm{xx}$ ($\Omega$) & $\Delta R_\mathrm{xx}/R_\mathrm{xx}$ (\%) & $\Delta R_\mathrm{xx}/R_\mathrm{xx}/j$ (arb.units)   \\
\hline
\hline
Ta/Co \cite{2} & $10^7$ & 574 & 0.011 &0.0019 & 1.3 \\
Pt/Co \cite{2} & $10^7$ & 176 & 0.0025 &0.0014 & 1 \\
GaMnAs heterostructure  \cite{1} & $7.5\times10^5$ & 1720 & 2 &0.12 & $1.1\times10^3$ \\
CBST/BST (this study)   & $5.0\times10^3$ $^*$ & 14000 & 57 &0.41 & $5.7\times10^5$ $^*$ \\
\end{tabular}
\end{table*}

To elucidate a possible origin of such a large UMR, we investigated angular dependence of the signal. Figures 2(b) and 2(c) show the in-plane magnetic-field directional dependence of normalized $\Delta R_\mathrm{xx}$ and $M_\mathrm{y}$ ($\propto \cos\varphi$); here definition of azimuth angle $\varphi$ is shown in Fig. 2(a). The $|\Delta R_\mathrm{xx}|$ is largest at $B||y$-axis ($\varphi=0^\circ,$ $180^\circ$ and $360^\circ$), scaling well with the $\cos \varphi$ dependence of $M_\mathrm{y}$. Figures 2(e) and 2(f) show the out-of-plane magnetic-field directional dependence of normalized $\Delta R_\mathrm{xx}$ (see Fig. 2(d)). Here, $M_\mathrm{y}$ and $M_\mathrm{z}$ are estimated from the variation of anomalous Hall effect. In accord with the in-plane case, the $|\Delta R_\mathrm{xx}|$ is largest at $B||y$-axis ($\theta=-90^\circ,$ $90^\circ$). It is noticeable, however, that the $\Delta R_\mathrm{xx}$ does not simply scale with $M_\mathrm{y}$. This is perhaps because the finite $M_\mathrm{z}$ component makes the Dirac dispersion massive, which effectively weakens the spin-momentum locking \cite{21}. To summarize, UMR emerges only when $M_\mathrm{y}$ component is finite. 
\begin{figure}
\centering
\includegraphics{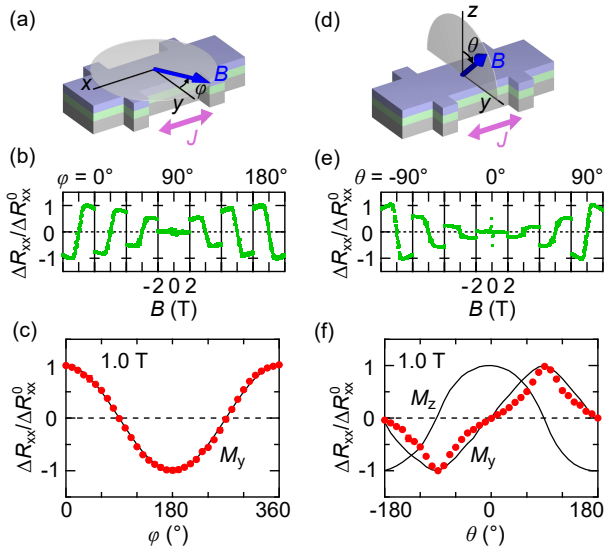}
\caption{(color online) (a) Schematic sample configuration for the measurement of in-plane magnetic field $\varphi$ dependence of $\Delta R_\mathrm{xx}$. $\varphi$ is measured from $y$-axis. (b), (c) Magnetic field and $\varphi$ dependence (by $30^\circ$ step) of normalized $\Delta R_\mathrm{xx}$, $\Delta R_\mathrm{xx}/\Delta R_\mathrm{xx}^0$. Here, $\Delta R_\mathrm{xx}^0$ is $\Delta R_\mathrm{xx}$ at $\varphi=0^\circ$ and 1.0 T. (d)-(f) The same as (a)-(c) for the out-of-plane magnetic field $\theta$ dependence. $\theta$ is measured from $z$-axis. Here, $\Delta R_\mathrm{xx}^0$ is $\Delta R_\mathrm{xx}$ at $\theta=90^\circ$ and 1.0 T. }
\label{F2}
\end{figure}

One possible origin of such nonlinear magnetoresistance might be an additional voltage caused by heat gradient along $z$-direction such as anomalous Nernst effect and spin Seebeck effect \cite{22,23,24}. In both processes, induced voltage would be expressed as $V_\mathrm{thermal} \propto M \times (\nabla T)_\mathrm{z}$ \cite{22,23,24}, so that finite $M_\mathrm{y}$ component might cause an additional voltage along the $x$-direction. However, we can safely exclude this possibility since the additional voltage should exhibit the same sign for the both heterostructures of CBST/BST/InP (Fig. 1(d)) and BST/CBST/InP (Fig. 1(g)) when InP works as a heat bath; this is inconsistent with the experimentally observed opposite sign shown in Figs. 1(f) and 1(i). Therefore, the origin of UMR should be explored in intrinsic scattering mechanisms related to electron spins. To clarify the microscopic origin, we studied the temperature dependence under higher magnetic field (Fig. 3(a)). UMR at low magnetic field decreases with increasing temperatures until it almost vanishes at around Curie temperature $T_\mathrm{C} \sim 24$ K \cite{SI}, confirming its close relevance to the ferromagnetic magnetization. This is also evident from the absence of UMR within the present experimental error in the single-layer BST film \cite{SI}. On the other hand, UMR is strongly suppressed at high magnetic field, meaning that it does not simply scale with $M_\mathrm{y}$. This indicates that the UMR in TI cannot be explained in terms of the GMR mechanism that was proposed for the case of ferromagnet/normal metal bilayers \cite{1,2,3}. Rather, such a field induced suppression of $|$$\Delta R_\mathrm{xx}$$|$ is reminiscent of the cases of Spin Seebeck effect \cite{23,24} and magnon Hall effect \cite{25}, in which the magnon population and hence the signal magnitude are suppressed by gap opening of spin wave (magnon) at higher field. This leads us to consider the scattering of surface Dirac electrons by magnons as a microscopic origin of UMR. 

\begin{figure}
\centering
\includegraphics{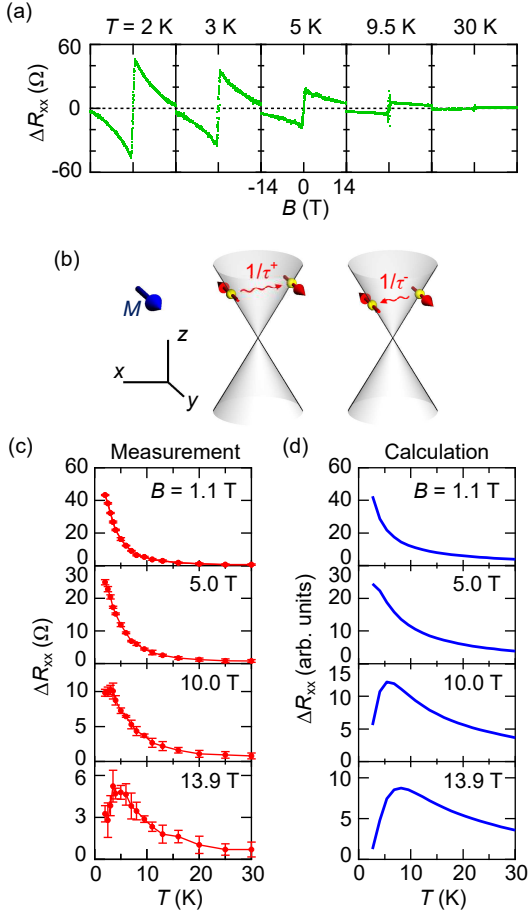}
\caption{(color online) (a) Magnetic field dependence of $\Delta R_\mathrm{xx}$ at various temperatures. (b) Schematic concepts of asymmetric scattering of spin-polarized surface Dirac electron by magnon. (c) Temperature dependence of $\Delta R_\mathrm{xx}$ under various magnetic fields. (d) Numerical calculation results of temperature dependent $\Delta R_\mathrm{xx}$ under various magnetic fields.}
\label{F3}
\end{figure}

With spin-momentum locking, conservation of angular momentum leads to the one-way scattering by magnon: Taking the quantization direction along $M||y$-axis, the angular momentum of magnon is $+1$ (note that spin angular momentum points opposite to $M$). Thus, as shown in Fig. 3(b), when electron with $s_\mathrm{y}=-1/2$ spin (left branch) is back-scattered to $s_\mathrm{y}=1/2$ (right branch), the electron absorbs magnon because of the conservation of angular momentum. On the other hand, when it goes from $s_\mathrm{y}=1/2$ to $s_\mathrm{y}=-1/2$, it emits magnon as a reverse process. Phenomena related to such a transfer of angular momentum between electron spin and magnetization has been widely recognized in the field of spintronics \cite{4}; for example, spin Seebeck effect \cite{22,23,24}, spin Peltier effect \cite{26}, spin pumping \cite{27,28} and spin torque ferromagnetic resonance \cite{5,6,13,14}. In such a scattering process by magnon, we can derive the formula of UMR by Boltzmann transport equation with the relaxation time approximation \cite{29} as follows (see Supplemental Material \cite{SI});
\begin{align}
\Delta R_\mathrm{xx} &
\propto j_\mathrm{x} \int {k_\mathrm{x}}
\left(-\frac{1}{\tau^+}+\frac{1}{\tau^-}\right)
\left(\frac{\partial ^2 f}{\partial {E}^2}\right),\\
\frac{1}{\tau^+} &
\propto\frac{1}{e^{\beta\hbar \omega}-1}\left({1-\frac{1}{e^{\beta(\hbar \omega+\hbar v_\mathrm{F}  k_\mathrm{x}-E_\mathrm{F}))}+1}}\right), \\
\frac{1}{\tau^-}&
\propto\left(\frac{1}{e^{\beta\hbar \omega}-1}+1\right)\left({1-\frac{1}{e^{\beta(-\hbar \omega+\hbar v_\mathrm{F} k_\mathrm{x}-E_\mathrm{F})}+1}}\right).
\end{align}
Here, $f$ is the Fermi distribution function and $\hbar \omega$ is the magnon energy with wavenumber $\sim 2k_\mathrm{F}$ ($k_\mathrm{F}$ : Fermi wavenumber). $\tau^+$ $(\tau^-)$ is relaxation time of magnon scattering from left(right) branch to right (left) one. The first factors of equations (3) and (4) are the probability of magnon absorption and emission, respectively, and the second ones are the probability that the final state of electron is unoccupied. Since $1/\tau^+$  and $1/\tau^-$  are not equal in general, equation (2) gives finite UMR in TI. Note that equation (2) is derived for the one-dimensional (1D) Dirac dispersion but this can be readily extended to the actual 2D case without essential change of the scheme \cite{SI}. 

Figures 3(c) and 3(d) display the comparison of experimental results of the temperature and magnetic-field dependence of UMR with the calculated ones based on the above model assuming the magnon energy of $g\mu_\mathrm{B}B$ \cite{30} with $g \sim 2$ for the localized Cr moment in CBST. Both results give qualitative consistency; the UMR monotonically increases with decreasing temperature at low magnetic field (1.1 T and 5.0 T), while the magnon gap as large as $\sim 20$ K opens at 13.9 T so that the UMR takes a peak structure around the temperature comparable with the magnon gap. Here, the deviation of numerical calculation from experimental result above 10 K originates from the breakdown of the adopted spin wave approximation \cite{SI} at temperatures close to $T_\mathrm{C} \sim 24$ K. This microscopic model helps us to understand why the UMR in TI is so large: One reason is the spin-momentum locking inherent in TI. Unlike the Rashba interface with two bands having opposite spin helicity, TI with single spin-momentum locking can accumulate spin efficiently without cancellation. Another factor is that TI with tuned $E_\mathrm{F}$ around the Dirac point can have a small Fermi momentum $k_\mathrm{F}$ lower than $\sim 500\ \mu \mathrm{m}^{-1}$ \cite{17,18}. Therefore, magnons with small wavenumber and low energy can dominantly contribute to electron scattering, which is easily populated even at low temperatures.

\begin{figure}
\centering
\includegraphics{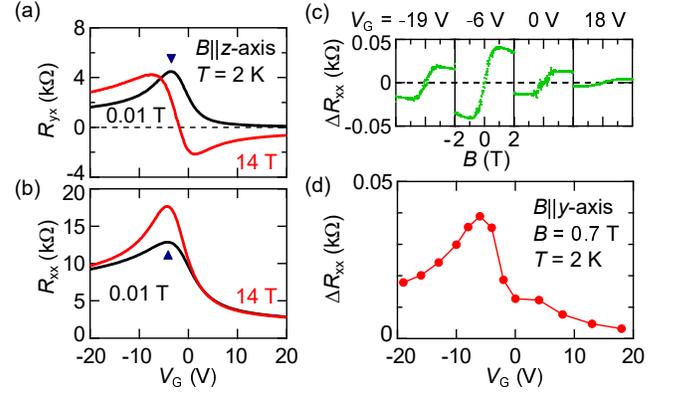}
\caption{(color online) (a), (b) Gate voltage $V_\mathrm{G}$ dependence of Hall resistance ($R_\mathrm{yx}$) and longitudinal resistance ($R_\mathrm{xx}$) under magnetic field $B$($||z$) of 0.01 T and 14 T at 2 K. (c), (d) Magnetic field and $V_\mathrm{G}$ dependence of $\Delta R_\mathrm{xx}$. The $\Delta R_\mathrm{xx}$ is taken at $B$($||y$)$ = 0.7$ T.}
\label{F4}
\end{figure} 

Finally, we discuss the $E_\mathrm{F}$ dependence of UMR in the field-effect transistor of TI heterostructure. Here, AlOx layer with a thickness of 30 nm was deposited as a top gate dielectric. Figures 4(a) and 4(b) show the gate voltage $V_\mathrm{G}$ dependence of $R_\mathrm{yx}$ and $R_\mathrm{xx}$ under $B||z$-axis at 0.01 T and 14 T. Anomalous Hall effect ($R_\mathrm{yx}$ measured at 0.01 T, shown by a black line in Fig. 4(a)) and $R_\mathrm{xx}$ (Fig. 4(b)) take maxima at around $-4$ V, indicating that $E_\mathrm{F}$ of the top surface state is tuned close to the Dirac point \cite{15}. The $V_\mathrm{G}$ dependence of UMR measured under $B||y$-axis is summarized in Figs. 4(c) and 4(d). To exclude the $V_\mathrm{G}$ dependence of relaxation time (denoted as $\tau^{(0)}$ in Supplemental Material), we plot $\Delta R_\mathrm{xx}$, not $\Delta R_\mathrm{xx}/R_\mathrm{xx}$, in Fig. 4(d) \cite{SI}. First, the sign of $\Delta R_\mathrm{xx}$ does not change with $V_\mathrm{G}$, $i.e.$ irrespective of $E_\mathrm{F}$ position in hole ($V_\mathrm{G} = -19$ V) and electron ($V_\mathrm{G}$ = 0 V and 18 V) doping regions. This can be understood by considering the scattering process for the hole side in the same way as shown in Fig. 3(b) for the electron side \cite{SI}. Moreover, the UMR is maximized at $E_\mathrm{F}$ being close to the Dirac point ($V_\mathrm{G} = -6$ V). As $k_\mathrm{F}$ decreases with $E_\mathrm{F}$ approaching the Dirac point, the wavenumber and energy of magnon contributing to the scattering process decrease so that related magnon population increases. This, in combination with the decrease of $k_\mathrm{F}$, results in the maximum $\Delta R_\mathrm{xx}$ and UMR with $E_\mathrm{F}$ around the Dirac point.

To summarize, we observed UMR in magnetic TI, which is shown to be several orders of magnitude larger than in other reported systems \cite{1,2,3}. The origin of UMR is identified to be the asymmetric scattering of electrons by magnons. Improvement of theoretical calculation \cite{SI} and understanding of the relationship with spin-orbit torque remain as future issues \cite{12,13,14}.\\

We thank T. Yokouchi, T. Ideue, Y. Okamura, N. Ogawa, S. Seki, K. Hamamoto and N. Nagaosa for fruitful discussions, and S. Shimizu for experimental support. This research was supported by the Japan Society for the Promotion of Science through the Funding Program for World-Leading Innovative R \& D on Science and Technology (FIRST Program) on ``Quantum Science on Strong Correlation" initiated by the Council for Science and Technology Policy and by JSPS Grant-in-Aid for Scientific Research(S) No. 24224009 and No. 24226002 from MEXT, Japan.

\pagebreak
\widetext
\def\baselinestretch{1.3}
\pagestyle{empty}
\pagestyle{plain}
\renewcommand{\bibnumfmt}[1]{[S#1]}

\newcommand{\BST }{(Bi$_{1-x}$Sb$_x$)$_2$Te$_3$}
\newcommand{\BSTy}{(Bi$_{1-y}$Sb$_y$)$_2$Te$_3$}
\newcommand{\CBST }{Cr$_x$(Bi$_{1-y}$Sb$_y$)$_{2-x}$Te$_3$}
\newcommand{\VBST }{V$_x$(Bi$_{1-y}$Sb$_y$)$_{2-x}$Te$_3$}
\newcommand{\CBSTBST}{Cr$_x$(Bi$_{1-y}$Sb$_y$)$_{2-x}$Te$_3$/(Bi$_{1-y}$Sb$_y$)$_2$Te$_3$}
\newcommand{\BSTCBST}{(Bi$_{1-y}$Sb$_y$)$_2$Te$_3$/Cr$_x$(Bi$_{1-y}$Sb$_y$)$_{2-x}$Te$_3$}

\renewcommand{\figurename}{Fig.}
\renewcommand{\thefigure}{S\arabic{figure}}

\newcounter{equationnum}
\setcounter{equationnum}{0}

\newcounter{secnum}
\setcounter{secnum}{0}

\newcounter{fignum}
\setcounter{fignum}{0}

\setcounter{equation}{0}
\setcounter{figure}{0}
\setcounter{table}{0}
\setcounter{page}{1}

\begin{center}
\textbf{Supplemental Material for\\\vspace{0.3cm}
\large  Large Unidirectional Magnetoresistance in a Magnetic Topological Insulator}
\end{center}

\stepcounter{secnum}
\section*{Supplemental Material S\thesecnum\ \textbar\ Device structure}
\begin{figure}[H]
\centering
\includegraphics{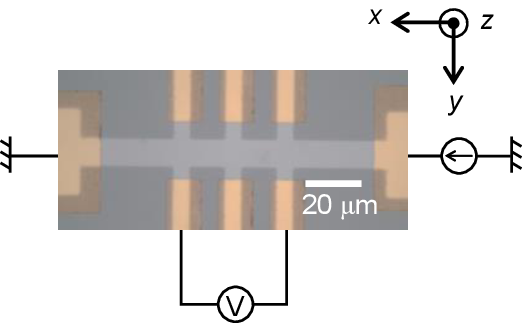}
\caption{top view of the device structure.}
\label{device_structure}
\end{figure}

Figure \ref{device_structure} shows the top view of the device structure. The size of the Hall bar is 10 $\mu$m in width and 36 $\mu$m in length, respectively. 

\stepcounter{secnum}
\section*{Supplemental Material S\thesecnum\ \textbar\ Magnetic anisotropy and band structure}
\begin{figure}[H]
\centering
\includegraphics{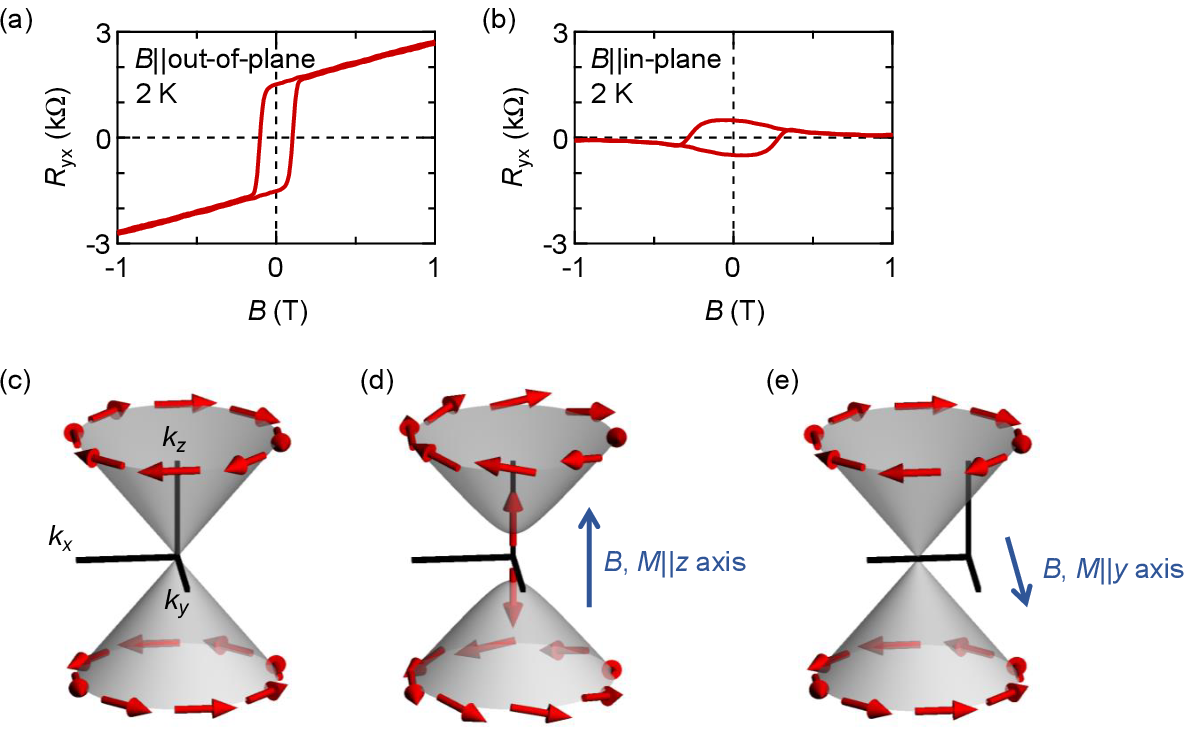}
\caption{(a) Magnetic field dependence of Hall resistance $R_{yx}$ with $B||$out-of-plane. (b) Same as (a) with $B||$in-plane. $R_{yx}$ is almost zero at and above 0.7 T, meaning $M||$in-plane. (c) Schematic diagram of the band structure without $B$ and $M$. (d) with $B,\ M||z$-axis. (e) with $B,\ M||y$-axis.}
\label{Magnetic_anisotropy}
\end{figure}

\CBST \ shows a perpendicular magnetic anisotropy. Thus, $M$ is pointing along $z$-direction at 0 T with the coercive field of $\sim$ 0.1 T (Fig. \ref{Magnetic_anisotropy}(a)). On the other hand, $M$ is directed to in-plane when the in-plane magnetic field of $\sim$ 0.7 T is applied (Fig. \ref{Magnetic_anisotropy}(b)). Because of the coupling between electron spin and localized moment, the Hamiltonian of surface electron is affected by the direction of magnetization $M$. The original Hamiltonian without $M$ is expressed as, 
\begin{equation}
H=\alpha(k_y\sigma_x-k_x\sigma_y),
	\stepcounter{equationnum}
	\tag{S\theequationnum}
\end{equation}
as displayed in Fig. \ref{Magnetic_anisotropy}(c). With the out-of-plane $M||z$, it turns to
\begin{equation}
H=\alpha(k_y\sigma_x-k_x\sigma_y)+m\sigma_z,
	\stepcounter{equationnum}
	\tag{S\theequationnum}
\end{equation}
so that the Dirac dispersion gets massive (Fig. \ref{Magnetic_anisotropy}(d)). On the other hand, with the in-plane $M||y$, it turns to
\begin{equation}
H=\alpha k_y\sigma_x+(m-\alpha k_x)\sigma_y,
	\stepcounter{equationnum}
	\tag{S\theequationnum}
\end{equation}
which shifts the gapless Dirac cone to the $k_x$-direction (Fig. \ref{Magnetic_anisotropy}(e)).

\stepcounter{secnum}
\section*{Supplemental Material S\thesecnum\ \textbar\ Estimation of heating by current injection}
\begin{figure}[H]
\centering
\includegraphics{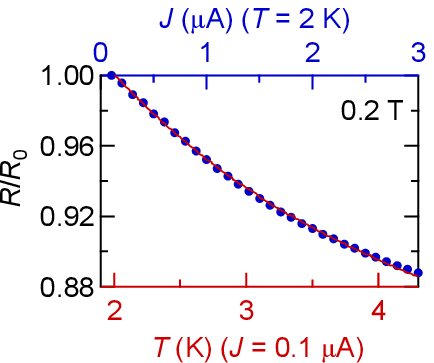}
\caption{The current dependence (at $T$ = 2 K) and temperature dependence (at $J$ = 0.1 $\mu$A) of normalized resistance. The measurements were done at 0.2 T with $B||z$-axis.}
\label{estimation_heating}
\end{figure}

Figure \ref{estimation_heating} shows the estimation of heating by current injection. The resistance is decreased by heating as we apply larger current. By comparing with temperature dependence of $R/R_0$ measured with minimal current (0.1 $\mu$A), the sample temperature is estimated to increase from 2 K to 2.7 K when applying 1 $\mu$A. Also, with 3 $\mu$A, it increases from 2 K to 4.3 K. For most measurements except for the current dependence, we applied $\pm$1 $\mu$A to get enough S/N ratio but to make the heating effect as small as possible. This heating effect, combined with the decrease of signal at high temperature (top panel of Fig. 3(c) in the main text), results in the breakdown of the linear relationship between $\Delta R_{xx}$ and $J$ at a higher current region in Fig. 1(k) in the main text.

\stepcounter{secnum}
\section*{Supplemental Material S\thesecnum\ \textbar\ Estimation of Curie temperature}
\begin{figure}[H]
\centering
\includegraphics{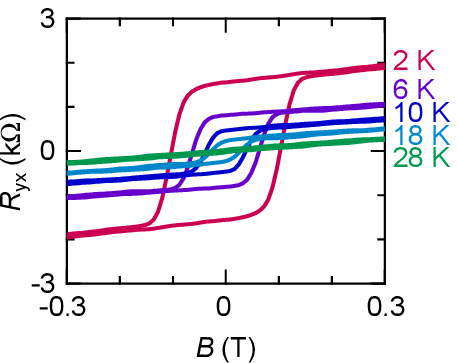}
\caption{Magnetic field $B||z$-axis dependence of Hall resistance $R_{yx}$ at selected temperatures.}
\label{arrott}
\end{figure}

In Fig. \ref{arrott}, we show the magnetic field dependence of Hall resistance. The anomalous Hall effect and the coercive field decreases with the increase of temperature. From the temperature dependence of the anomalous Hall effect, we can estimate the $T_\mathrm{C}$ as $\sim$ 24 K.

\stepcounter{secnum}
\section*{Supplemental Material S\thesecnum\ \textbar\ Structure dependence of UMR}

Figure R5 shows the structure dependence of UMR for CBST/BST, BST/CBST, single-layer CBST and single-layer BST. Here, CBST/BST and BST/CBST are taken from Figs. 1(f) and 1(i) in the main text. In single-layer CBST, $\Delta R_{xx}$ takes a small finite value. Here, although the overall shape of the magnetic field dependence is the same as that in CBST/BST, the absolute value of the signal is about 10 times smaller. Since Cr is distributed over the whole film in CBST, both the top and bottom surfaces would exhibit finite UMR with an opposite sign, leading to the cancellation of the signal. Here, the observed smaller but finite signal probably originates from the difference in those environments. As for single-layer BST, on the other hand, no UMR signal is observed within the range of measurement error. This reconfirms the magnetic origin of UMR.

\begin{figure}[H]
\centering
\includegraphics{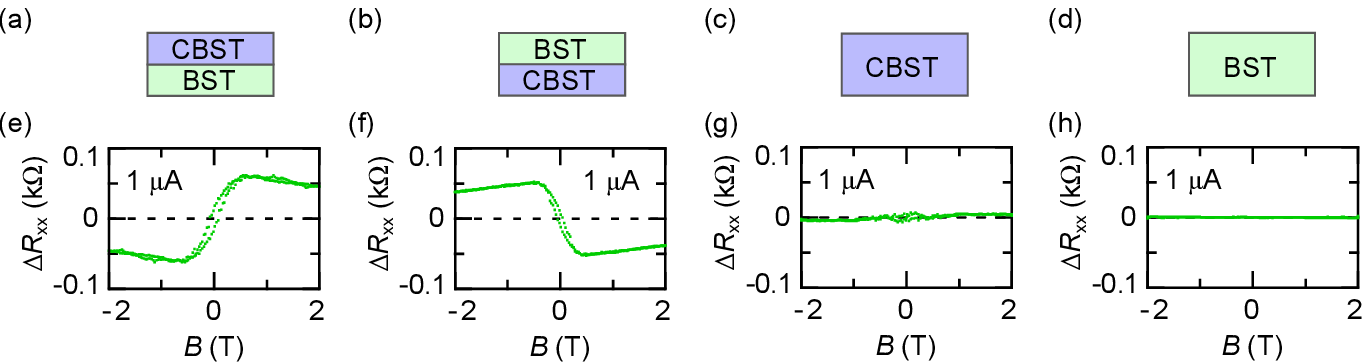}
\caption{(a)-(d) Schematic sample configurations for structure dependent characterization of UMR signal; CBST/BST (a), BST/CBST (b), CBST (c) and BST (d). The total thickness of each sample is fixed to 8 nm. (e)-(h) In-plane magnetic field dependence of $\Delta R_{xx}$ at $T$ = 2 K.}
\label{structure_dep}
\end{figure}

\stepcounter{secnum}
\section*{Supplemental Material S\thesecnum\ \textbar\ Derivation of nonlinear conduction}
Here, we derive the expression for the nonlinear Boltzmann transport equation using relaxation time approximation. When $\bm{E}$ is large enough, we need to discuss the nonlinear resistance beyond linear response $\bm{J}=\bm{\sigma}\bm{E}$.

First, we derive the relationship between $\bm{J}$ and $\bm{E}$ [S1]. We define the distribution function of electron as $g(\bm{r},\bm{k},t)=g(\bm{k})$. Here, we assume time-independent uniform electric field so that $g$ is dependent only on $\bm{k}$. When $\bm{E}=0$ (equilibrium), the distribution function is expressed by Fermi distribution function,
\begin{equation}
g^0(\bm{k})=f(\bm{k})=\frac{1}{e^{(\epsilon(k)-\mu)/k_\mathrm{B}T}+1}.
\label{distribution_eq}
	\stepcounter{equationnum}
	\tag{S\theequationnum}
\end{equation}
We first consider the scattering of electron by impurity, phonon, electron-electron, magnon $etc$. In the relaxation time approximation, the electron is assumed to relax to equilibrium within the typical timescale of $\tau$ as follows,
\begin{equation}
\left( \frac{dg(\bm{k})}{dt}\right)_\mathrm{coll}=-\frac{g(\bm{k})-g^0(\bm{k})}{\tau(\bm{k})}.
\label{collision}
	\stepcounter{equationnum}
	\tag{S\theequationnum}
\end{equation}

Next, assuming equation (\ref{collision}), we derive the nonlinear distribution function. Boltzmann equation is expressed as,
\begin{equation}
\frac{\partial g}{\partial t}+\bm{v}\cdot \frac{\partial g}{\partial \bm{r}}+\bm{F}\cdot\frac{1}{\hbar}\frac{\partial g}{\partial \bm{k}}=\left( \frac{dg(\bm{k})}{dt}\right)_\mathrm{coll}.
\label{boltzman}
	\stepcounter{equationnum}
	\tag{S\theequationnum}
\end{equation}
In the steady state under uniform electric field,
\begin{equation}
\frac{\partial g}{\partial t}=0,\ \frac{\partial g}{\partial \bm{r}}=0,\ \bm{F}=-e\bm{E},
	\stepcounter{equationnum}
	\tag{S\theequationnum}
\end{equation}
so that the equation (\ref{boltzman}) is expressed as,
\begin{equation}
-\frac{e\bm{E}}{\hbar}\cdot\frac{\partial g}{\partial \bm{k}}=-\frac{g-g^0}{\tau}.
	\stepcounter{equationnum}
	\tag{S\theequationnum}
\end{equation}
Namely,
\begin{equation}
g=f+\frac{e\bm{E}\tau}{\hbar}\cdot\frac{\partial g}{\partial \bm{k}}.
\label{g_master_eq}
	\stepcounter{equationnum}
	\tag{S\theequationnum}
\end{equation}
In the following, we consider the case of $\bm{E}=(E_x,0,0)$ for simplicity. The linear approximation leads to the well-known relation [S1],
\begin{align}
j_x&=\sigma_{xx}^{(1)}E_x,
	\stepcounter{equationnum}
	\tag{S\theequationnum} \\
\sigma_{xx}^{(1)}=e^2\int \frac{d\bm{k}}{4\pi^3}\tau(\bm{k})&v_x(\bm{k})v_x(\bm{k})\left(-\frac{\partial f}{\partial {\epsilon}}\right).
	\stepcounter{equationnum}
	\tag{S\theequationnum}
\label{sigma_xx}
\end{align}Expanding the formula up to $E^2$ term, $i.e.$,
\begin{align}
g&=f+\frac{e{E}\tau}{\hbar}\frac{\partial }{\partial {k_x}}\left( f+\frac{e{E}\tau}{\hbar}\frac{\partial f}{\partial {k_x}}\right)
	\stepcounter{equationnum}
	\tag{S\theequationnum} \\
&=f+\frac{e{E}\tau}{\hbar}\frac{\partial f}{\partial {k_x}}+\left(\frac{e{E}\tau}{\hbar}\right)^2\left(\frac{\partial ^2 f}{\partial {k_x}^2}\right),
	\stepcounter{equationnum}
	\tag{S\theequationnum}
\end{align}
we obtain the non-linear conductivity,
\begin{align}
j_x&=\sigma_{xx}^{(1)}E_x+\sigma_{xx}^{(2)}E_x^2, 
\label{nonlinar_equation}
	\stepcounter{equationnum}
	\tag{S\theequationnum}\\
\sigma_{xx}^{(2)}=-\frac{e^3}{\hbar^2}&\int \frac{d\bm{k}}{4\pi^3}\left(\tau(\bm{k})\right)^2 v_x(\bm{k}) \left(\frac{\partial ^2 f}{\partial {k_x}^2}\right).
	\stepcounter{equationnum}
	\tag{S\theequationnum}
\label{sigma_xx2}
\end{align}

In order to make comparison between the theory and the measured resistance, we discuss the relationship between $\Delta R_{xx}$ and $\sigma_{xx}^{(2)}$. To the first order approximation of equation (\ref{nonlinar_equation}), we get
\begin{equation}
V_x=\frac{A}{l}E_x=\frac{A}{l}\left(\frac{1}{\sigma_{xx}^{(1)}}j_x-\frac{\sigma_{xx}^{(2)}}{(\sigma_{xx}^{(1)})^3}j_x^2 \right).
	\stepcounter{equationnum}
	\tag{S\theequationnum}
\end{equation}
Here, $A$ and $l$ is the cross-sectional area and length of the Hall bar, respectively. Thus, when we excite plus or minus dc current, the measured resistance is,
\begin{align}
R_{xx}^+&=\frac{1}{l}\left(\frac{1}{\sigma_{xx}^{(1)}}-\frac{\sigma_{xx}^{(2)}}{(\sigma_{xx}^{(1)})^3}j_x\right),
	\stepcounter{equationnum}
	\tag{S\theequationnum}\\
R_{xx}^-&=\frac{1}{l}\left(\frac{1}{\sigma_{xx}^{(1)}}+\frac{\sigma_{xx}^{(2)}}{(\sigma_{xx}^{(1)})^3}j_x\right), 
	\stepcounter{equationnum}
	\tag{S\theequationnum}\\
R_{xx}&=\frac{R_{xx}^++R_{xx}^-}{2}=\frac{1}{l}\left(\frac{1}{\sigma_{xx}^{(1)}}\right), 
	\stepcounter{equationnum}
	\tag{S\theequationnum}\\
\Delta R_{xx}&={R_{xx}^+-R_{xx}^-}=-\frac{1}{l}\left(\frac{2\sigma_{xx}^{(2)}}{(\sigma_{xx}^{(1)})^3}j_x\right). 
\label{delta_Rxx}
	\stepcounter{equationnum}
	\tag{S\theequationnum}
\end{align}
In the following, we discuss the microscopic origin of $\Delta R_{xx}$ and UMR.

\stepcounter{secnum}
\section*{Supplemental Material S\thesecnum\ \textbar\ Microscopic origin of UMR}
Here, we discuss the microscopic origin of $\sigma_{xx}^{(2)}$ in magnetic TI. For simplicity, we first consider the 1D Dirac dispersion. According to equation (\ref{sigma_xx2}) the origin of $\sigma_{xx}^{(2)}$ can be divided into,
\begin{enumerate}
\renewcommand{\labelenumi}{\roman{enumi}).}
\item asymmetry between $+\bm{k}$ and $-\bm{k}$ in $v_{x}(\bm{k})\left(\frac{\partial^2 f}{\partial k_x^2}\right)$, 
\item asymmetry between $+\bm{k}$ and $-\bm{k}$ in $(\tau(\bm{k}))^2$.
\end{enumerate}
In completely $k$-linear Dirac dispersion, the Dirac point and the whole dispersion shift to the $k_x$-direction under magnetic field and/or magnetization along the $y$-direction so that $v_x$ is unchanged (Fig. S2e). However, in the surface state of \BST, $k^2$ term and hexagonal warping ($k^3$ term) actually exist in addition to the $k$-linear term [S2]. Therefore, because of the contribution from i), $\sigma_{xx}^{(2)}$ can be finite and is expected to increase as the in-plane component of magnetization $M_y$ increases. In reality, as shown in Fig. 3(a) in the main text, $\Delta R_{xx}$ decreases as a function of the magnetic field ($>$ 1 T). Therefore, the relative contribution from i) seems to be small in the present case.

\begin{figure}[H]
\centering
\includegraphics{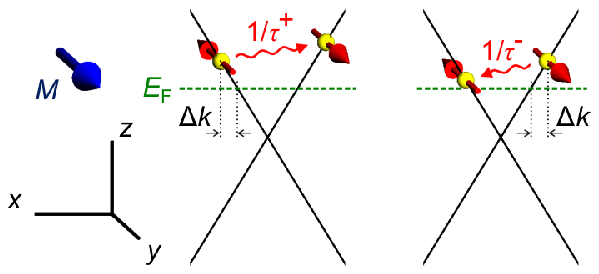}
\caption{Schematic pictures of asymmetric scattering of electron by magnon, where $E_\mathrm{F}$ is at the electron side. The $\Delta k$ is measured from $k_\mathrm{F}$ (Fermi wavenumber).}
\label{1D_diagram}
\end{figure}

Next we discuss the contribution from ii). From the temperature dependence in Fig. 3(a) in the main text, we already know that UMR is related to magnetism. Therefore, it is natural for us to consider the scattering of electrons by magnetic excitations, or magnons. Here, we consider the interaction between the surface conduction electron composed of Bi, Sb and Te $p$ orbital and the localized spin composed of Cr $d$ orbital. When $M$ is pointing along $y$ direction, the localized spin is pointing in the $-y$ direction. Therefore, the angular momentum of magnon is +1. Here, by the conservation of angular momentum, the interaction Hamiltonian $\mathcal{H}'$ is written as,
\begin{align}
\mathcal{H}'&\propto\sum_i \left( c_{i,\uparrow}^\dagger c_{i,\downarrow} b_i  + c_{i,\downarrow}^\dagger c_{i,\uparrow}b_i^\dagger \right) 
\label{interaction_H_local}
	\stepcounter{equationnum}
	\tag{S\theequationnum}\\
 &=\sum_{k,q} \left(c_{k+q,\uparrow}^\dagger c_{k,\downarrow} b_q  + c_{k-q,\downarrow}^\dagger c_{k,\uparrow}b_q^\dagger \right).
\label{interaction_FT}
	\stepcounter{equationnum}
	\tag{S\theequationnum}
\end{align}
Here, $b^\dagger$ ($b$) and $c^\dagger$ ($c$) are a creation (annihilation) operator of magnon and surface Dirac electron, respectively. The equation (\ref{interaction_H_local}) means that the interaction at sites $i$ can have only two types of process. One process is $c_{i,\uparrow}^\dagger c_{i,\downarrow} b_i $ meaning that electron spin is changed from $\downarrow (s_y=-\frac{1}{2})$ to $\uparrow (s_y=\frac{1}{2})$ by the absorption of magnon. This is because of the conservation of angular momentum; $-\frac{1}{2}+(+1)=\frac{1}{2}$ (note that the angular momentum of magnon is $+1$).
Another process is  $ c_{i,\downarrow}^\dagger c_{i,\uparrow}b_i^\dagger$ meaning that electron spin is reversed from $\uparrow (s_y=\frac{1}{2})$ to $\downarrow (s_y=-\frac{1}{2})$ by the emission of magnon ($\frac{1}{2}-(+1)=-\frac{1}{2}$). Because of the spin-momentum locking of TI, the electron absorbs magnon to move from left branch to right one, while it emits magnon to move from right branch to left one. Here, the related magnon wavenumber is about 2$k_\mathrm{F}$ because of the conservation of momentum. These processes are schematically shown in Fig. \ref{1D_diagram}. Because of such asymmetry in inelastic scattering by magnon, we naively expect asymmetry between $+\bm{k}$ and $-\bm{k}$ in $(\tau(\bm{k}))^2$, which leads to the UMR.

\stepcounter{secnum}
\section*{Supplemental Material S\thesecnum\ \textbar\ Calculation of UMR in 1D Dirac dispersion}
In the following, we derive the formula for $\Delta R_{xx}$ in 1D Dirac dispersion for the qualitative understanding of the magnon scattering process. There are many sources of scattering in crystal such as impurity, phonon, electron-electron scattering. We assume that magnon scattering and these processes are completely independent of each other, then the relaxation time is written as
\begin{equation}
\frac{1}{\tau}=\frac{1}{\tau^{(0)}}+\frac{1}{\tau_\mathrm{mag}}.
	\stepcounter{equationnum}
	\tag{S\theequationnum}
\label{Matthiessen}
\end{equation}
Here, $\tau^{(0)}$ is nonmagnetic scattering and $\tau_\mathrm{mag}$ is scattering by magnon. Since the impurity scattering is dominant, $\tau^{(0)}\ll \tau_\mathrm{mag}$ hold true in this case. Therefore, to the first order approximation, the relaxation time is expressed as,
\begin{equation}
{\tau}=\tau^{(0)}-\frac{(\tau^{(0)})^2}{\tau_\mathrm{mag}}.
	\stepcounter{equationnum}
	\tag{S\theequationnum}
\end{equation}
By substituting this to equation (\ref{sigma_xx2}), we obtain
\begin{align}
\sigma_{xx}^{(2)}&=-\frac{e^3}{\hbar^2}\int \frac{d\bm{k}}{4\pi^3}\left(\tau(\bm{k})\right)^2 v_x(\bm{k}) \left(\frac{\partial ^2 f}{\partial {k_x}^2}\right)
	\stepcounter{equationnum}
	\tag{S\theequationnum} \\
 &=-\frac{e^3}{\hbar^2}\int \frac{d\bm{k}}{4\pi^3}\left((\tau^{(0)})^2-\frac{2(\tau^{(0)})^3}{\tau_\mathrm{mag}(\bm{k})}\right)
v_x(\bm{k}) \left(\frac{\partial ^2 f}{\partial {k_x}^2}\right) 
\label{sigma_xx2_mag_v2}
	\stepcounter{equationnum}
	\tag{S\theequationnum}\\
&=\frac{2(\tau^{(0)})^3e^3}{\hbar^2}\int \frac{d\bm{k}}{4\pi^3}
\frac{1}{\tau_\mathrm{mag}(\bm{k})}
v_x(\bm{k}) \left(\frac{\partial ^2 f}{\partial {k_x}^2}\right). 
	\stepcounter{equationnum}
	\tag{S\theequationnum}
\label{sigma_xx2_mag}
\end{align}
Here, $\tau_\mathrm{mag}$ can be expressed as
\begin{align}
\frac{1}{\tau_\mathrm{mag}(\bm{k})}=&\int \frac{d\bm{k}'}{(2\pi)^3}W_\mathrm{mag}(\bm{k}'|\bm{k})(1-g(\bm{k}')), \label{tau_mag}
	\stepcounter{equationnum}
	\tag{S\theequationnum}
\\
W_\mathrm{mag}(\bm{k}'|\bm{k})=&W_\mathrm{abs}(\bm{k}',\sigma';n_{-\bm{k}+\bm{k}'} -1|\bm{k},\sigma;n_{-\bm{k}+\bm{k}'})
	\stepcounter{equationnum}
	\tag{S\theequationnum}\\
&+W_\mathrm{emit}(\bm{k}',\sigma';n_{\bm{k}-\bm{k}'} +1|\bm{k},\sigma;n_{\bm{k}-\bm{k}'}). 
	\stepcounter{equationnum}
	\tag{S\theequationnum}\nonumber
\end{align}
Here, $W_\mathrm{abs}$ and $W_\mathrm{emit}$ are scattering probability for the magnon absorption and emission process and $1-g(\bm{k}')$ is the probability that the final destination of electron is unoccupied. $\bm{k},\sigma$ represent the electron wavenumber and spin, $n_{\bm{k}}$ corresponds to the number of magnon, respectively. Here, scattering probability is represented as,

\begin{align}
W_\mathrm{abs}(\bm{k}',\sigma';n_{-\bm{k}+\bm{k}'} -1|\bm{k},\sigma;n_{-\bm{k}+\bm{k}'})&=\frac{2\pi}{\hbar}|\langle \bm{k}',\sigma';n_{-\bm{k}+\bm{k}'} -1|\mathcal{H}'|\bm{k},\sigma;n_{-\bm{k}+\bm{k}'}\rangle|^2\delta(\epsilon_{\bm{k}'}-\epsilon_{\bm{k}}-\hbar\omega_{-\bm{k}+\bm{k}'})   
	\stepcounter{equationnum}
	\tag{S\theequationnum} \\
&=\frac{2\pi}{\hbar}n_{-\bm{k}+\bm{k}'}|\langle\sigma'|c_{\uparrow}^\dagger c_{\downarrow}|\sigma \rangle |^2 \delta(\epsilon_{\bm{k}'}-\epsilon_{\bm{k}}-\hbar\omega_{-\bm{k}+\bm{k}'}). 
	\stepcounter{equationnum}
	\tag{S\theequationnum}
\label{w_abs}\\
W_\mathrm{emit}(\bm{k}',\sigma';n_{\bm{k}-\bm{k}'} +1|\bm{k},\sigma;n_{\bm{k}-\bm{k}'})&=\frac{2\pi}{\hbar}|\langle \bm{k}',\sigma';n_{\bm{k}-\bm{k}'} +1|\mathcal{H}'|\bm{k},\sigma;n_{\bm{k}-\bm{k}'}\rangle|^2\delta(\epsilon_{\bm{k}'}-\epsilon_{\bm{k}}+\hbar\omega_{\bm{k}-\bm{k}'})  
	\stepcounter{equationnum}
	\tag{S\theequationnum}
\\
&=\frac{2\pi}{\hbar}(n_{\bm{k}-\bm{k}'}+1)|\langle\sigma'|c_{\downarrow}^\dagger c_{\uparrow}|\sigma \rangle |^2 \delta(\epsilon_{\bm{k}'}-\epsilon_{\bm{k}}+\hbar\omega_{\bm{k}-\bm{k}'}), 
	\stepcounter{equationnum}
	\tag{S\theequationnum}
\label{w_emit}
\end{align}

Using equation (\ref{tau_mag}), we calculate $\tau_\mathrm{mag}^+$, scattering of the electron from the left branch (displaced from $k_\mathrm{F}$ by $\Delta k$) to the right branch,
\begin{equation}
\frac{1}{\tau_\mathrm{mag}^+(\Delta k)}\propto\frac{1}{e^{\beta\hbar \omega}-1}\left({1-\frac{1}{e^{\beta(\hbar \omega+\hbar v_\mathrm{F} \Delta k)}+1}}\right).
	\stepcounter{equationnum}
	\tag{S\theequationnum}
\label{mag+}
\end{equation}
Here, $\hbar{\omega}$ is an energy of magnon with $2k_\mathrm{F}$ wavenumber. The first factor is the Bose factor corresponding to magnon population and the second factor the probability that the final destination is unoccupied. In a similar way, the relaxation time from right to left is expressed as,
\begin{equation}
\frac{1}{\tau_\mathrm{mag}^-(\Delta k)}\propto\left(\frac{1}{e^{\beta\hbar \omega}-1}+1\right)\left({1-\frac{1}{e^{\beta(-\hbar \omega+\hbar v_\mathrm{F} \Delta k)}+1}}\right).
	\stepcounter{equationnum}
	\tag{S\theequationnum}
\label{mag-}
\end{equation}
Using these and equation (\ref{sigma_xx2_mag}), we finally obtain the expression for nonlinear conduction in 1D Dirac dispersion,
\begin{equation}
\sigma_{xx}^{(2)}\propto (\tau^{(0)})^3 \int {d\Delta{k}}\left(\frac{1}{\tau_\mathrm{mag}^+(\Delta k)}-\frac{1}{\tau_\mathrm{mag}^-(\Delta k)}\right)\left(\frac{\partial ^2 f}{\partial {k_x}^2}(\Delta{k})\right).
\label{1D_sigma_xx2}
	\stepcounter{equationnum}
	\tag{S\theequationnum}
\end{equation}
Since $E=v_\mathrm{F}k_x$, in 1D Dirac dispersion,
\begin{equation}
\frac{\partial ^2 f}{\partial {k_x}^2}=v_\mathrm{F}^2\frac{\partial ^2 f}{\partial {E}^2}.
	\stepcounter{equationnum}
	\tag{S\theequationnum}
\end{equation}
Also, using equation (\ref{sigma_xx}) we obtain,
\begin{equation}
\sigma_{xx}^{(1)}\propto \tau^{(0)}.
	\stepcounter{equationnum}
	\tag{S\theequationnum}
\end{equation}
Therefore, from equation (\ref{delta_Rxx}) we get the expression for UMR in 1D Dirac dispersion,
\begin{equation}
\Delta R_{xx} \propto j_x \int {d\Delta{k}}\left(-\frac{1}{\tau_\mathrm{mag}^+(\Delta k)}+\frac{1}{\tau_\mathrm{mag}^-(\Delta k)}\right)\left(\frac{\partial ^2 f}{\partial {E}^2}(\Delta{k})\right),
\label{1D_UMR}
	\stepcounter{equationnum}
	\tag{S\theequationnum}
\end{equation}
which is the equation (2) in the main text.

\begin{figure}[H]
\centering
\includegraphics{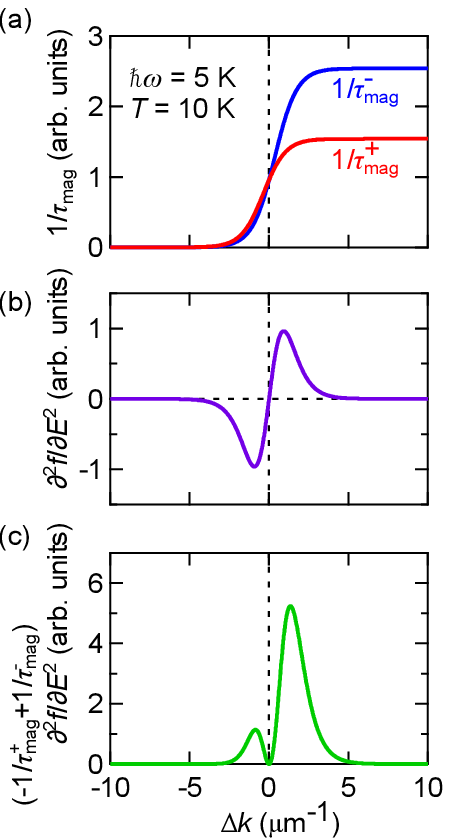}
\caption{(a) $\Delta k$ dependence of $1/\tau_\mathrm{mag}^-$ and $1/\tau_\mathrm{mag}^+$. (b) $\Delta k$ dependence of ${\partial ^2 f}/{\partial {E}^2}$. (c) $\Delta k$ dependence of the integrand of $\Delta R_{xx}$. The calculation is done at $\hbar\omega=5$ K and $T=10$ K.}
\label{calculation_1D}
\end{figure}

In Fig. \ref{calculation_1D}, we show the calculation result of equation (\ref{1D_UMR}). We can see that $1/\tau_\mathrm{mag}^-$ and $1/\tau_\mathrm{mag}^+$ show clear deviation. The important point is that when $\Delta k$ is positive, $1/\tau_\mathrm{mag}^- > 1/\tau_\mathrm{mag}^+$ and when it is negative, $1/\tau_\mathrm{mag}^- < 1/\tau_\mathrm{mag}^+$. Similarly, ${\partial ^2 f}/{\partial {E}^2}$ also changes its sign depending on the sign of $\Delta k$. Therefore, the integrand of equation (\ref{1D_UMR}) is positive for all $\Delta k$ so that we get finite $\Delta R_{xx}$.

We can derive exactly the same formula (\ref{1D_UMR}) for the hole side in the case when $E_\mathrm{F}$ position is at the hole side (Fig. S7). Therefore, $\Delta R_{xx}$ gives the same sign for the electron and the hole side. Note that the above treatment of magnetic excitation is based on the spin wave approximation and hence that the theory is only applicable at low enough temperatures as compared with $T_\mathrm{C}$ [S1, S3]. Nevertheless, $\Delta R_{xx}$ is expected to diminish around and above $T_\mathrm{C}$ with vanishing $M$, in accord with the experimental result (Fig. 3(c) in the main text).

\begin{figure}[H]
\centering
\includegraphics{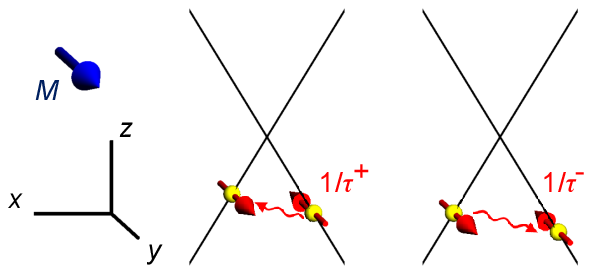}
\caption{Schematic pictures of asymmetric scattering at the hole side.}
\label{1D_diagram_hole}
\end{figure}

\stepcounter{secnum}
\section*{Supplemental Material S\thesecnum\ \textbar\ Calculation of UMR in 2D Dirac dispersion}
\begin{figure}[H]
\centering
\includegraphics{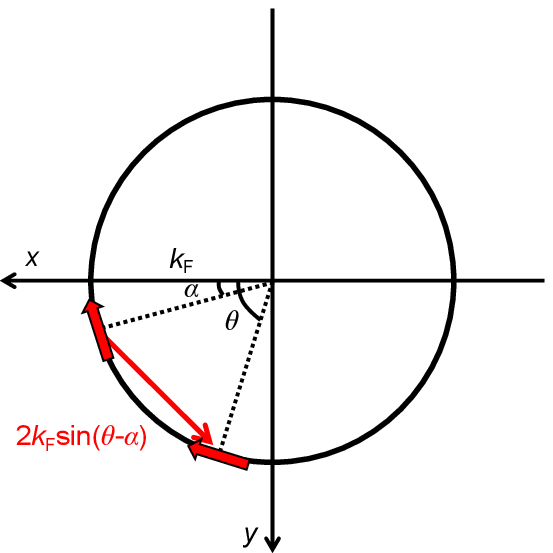}
\caption{Top view of the magnon scattering in 2D Dirac dispersion.}
\label{calculation_2D}
\end{figure}

Here, we expand the calculation of UMR in the 1D case to the 2D case. For simplicity, here we consider the linear dispersion again. The important difference from the 1D case is that the direction of spin is not fixed to the $y$-direction. We consider the scattering from the position $\alpha$ to $\theta$ as shown in Fig. \ref{calculation_2D}. Here, spin eigenfunction at $\alpha$ is, 
\begin{equation}
|\alpha\rangle=\sin\frac{\alpha}{2}|\uparrow\rangle+\cos\frac{\alpha}{2}|\downarrow\rangle.
	\stepcounter{equationnum}
	\tag{S\theequationnum}
\end{equation}
Hence, the factors in equations (\ref{w_emit}) and (\ref{w_abs}) are expressed as,
\begin{align}
|\langle\theta|c_{\uparrow}^\dagger c_{\downarrow}|\alpha\rangle|^2&=\cos^2\frac{\alpha}{2}\times\sin^2\frac{\theta}{2}, 
	\stepcounter{equationnum}
	\tag{S\theequationnum}\\
|\langle\theta|c_{\downarrow}^\dagger c_{\uparrow}|\alpha\rangle|^2&=\sin^2\frac{\alpha}{2}\times\cos^2\frac{\theta}{2}. 
	\stepcounter{equationnum}
	\tag{S\theequationnum}
\end{align}

Therefore, when we define the integrand of equation (\ref{tau_mag}) as $V_\mathrm{mag}$,
\begin{align}
V_\mathrm{mag}(\theta,\alpha,\Delta{k})&= \cos^2\frac{\alpha}{2}\sin^2\frac{\theta}{2}{V_\mathrm{mag}^+(\alpha,\theta,\Delta{k})}  +
 \sin^2\frac{\alpha}{2}\cos^2\frac{\theta}{2}{V_\mathrm{mag}^-(\alpha,\theta,\Delta{k})}, 
	\stepcounter{equationnum}
	\tag{S\theequationnum}\\
{V_\mathrm{mag}^+(\theta,\alpha,\Delta{k})}&\propto\frac{1}{e^{\beta\hbar \omega}-1}\left({1-\frac{1}{e^{\beta(\hbar \omega+\hbar v_\mathrm{F} \Delta k)}+1}}\right),
	\stepcounter{equationnum}
	\tag{S\theequationnum} \\
{V_\mathrm{mag}^-(\theta,\alpha,\Delta{k})}&\propto\left(\frac{1}{e^{\beta\hbar \omega}-1}+1\right)\left({1-\frac{1}{e^{\beta(-\hbar \omega+\hbar v_\mathrm{F} \Delta k)}+1}}\right).
	\stepcounter{equationnum}
	\tag{S\theequationnum}
\end{align}
Here, $\hbar \omega$ corresponds to the magnon energy with $2k_\mathrm{F} \sin(\theta-\alpha)$ wavenumber. Using equation (\ref{tau_mag}), we get
\begin{align}
\frac{1}{\tau_\mathrm{mag}(\alpha,\Delta{k})}&=\int \frac{d\bm{k}'}{(2\pi)^3}V_\mathrm{mag}(\theta,\alpha,\Delta{k}) 
	\stepcounter{equationnum}
	\tag{S\theequationnum}\\
&=\int \frac{d{k}'}{(2\pi)^3} \int d\theta k' V_\mathrm{mag}(\theta,\alpha,\Delta{k})
	\stepcounter{equationnum}
	\tag{S\theequationnum}\\
&=\frac{k_{F}}{{(2\pi)^3}} \int d\theta  V_\mathrm{mag}(\theta,\alpha,\Delta{k}).
	\stepcounter{equationnum}
	\tag{S\theequationnum}
\end{align}
As for $v_x(\bm{k})$,
\begin{equation}
v_x(\alpha,\Delta{k})=v_\mathrm{F}\cos \alpha. 
	\stepcounter{equationnum}
	\tag{S\theequationnum}
\end{equation}
As for $\frac{\partial ^2 f}{\partial {k_x}^2}$,
\begin{align}
\frac{\partial ^2 f}{\partial {k_x}^2}(\alpha,\Delta{k})&=\frac{\beta \hbar v_\mathrm{F}}{k_\mathrm{F}}\left( \frac{-P}{(P+1)^2}\right)\sin^2 \alpha+({\beta \hbar v_\mathrm{F}})^2 \left( \frac{P(P-1)}{(P+1)^3}\right)\cos^2 \alpha 
	\stepcounter{equationnum}
	\tag{S\theequationnum} \\
&\simeq({\beta \hbar v_\mathrm{F}})^2 \left( \frac{P(P-1)}{(P+1)^3}\right)\cos^2 \alpha, \label{diff_2D}
	\stepcounter{equationnum}
	\tag{S\theequationnum}\\
P&=e^{\beta \hbar v_\mathrm{F} \Delta k}.
	\stepcounter{equationnum}
	\tag{S\theequationnum}
\end{align}
Here, in equation (\ref{diff_2D}), we can ignore the first term since $\beta \hbar v_\mathrm{F} k_\mathrm{F}\gg 1$ except for the immediate vicinity of the Dirac point. Summarizing these, we can calculate $\sigma_{xx}^{(2)}$ from equation (\ref{sigma_xx2_mag});
\begin{align}
\sigma_{xx}^{(2)}&\propto (\tau^{(0)})^3{k_{F}}\int d\bm{k}\int d\theta  \left[V_\mathrm{mag}(\theta,\alpha,\Delta{k})\left(\frac{\partial ^2 f}{\partial {k_x}^2} \right) v_\mathrm{F}\cos \alpha\right]
	\stepcounter{equationnum}
	\tag{S\theequationnum}\\
&=(\tau^{(0)})^3{k_{F}}^2\int d\Delta{k}\int d\alpha\int d\theta \left[ V_\mathrm{mag}(\theta,\alpha,\Delta{k})\left(\frac{\partial ^2 f}{\partial {k_x}^2} \right) v_\mathrm{F}\cos \alpha \right].
	\stepcounter{equationnum}
	\tag{S\theequationnum}
\end{align}
Here, we ignored the anisotropy of $\tau^{(0)}$. Also, using equation (\ref{sigma_xx}) we obtain,
\begin{equation}
\sigma_{xx}^{(1)}\propto \tau^{(0)}k_\mathrm{F}.
	\stepcounter{equationnum}
	\tag{S\theequationnum}
\end{equation}
Therefore, from equation (\ref{delta_Rxx}) we finally obtain the expression for the UMR of 2D Dirac dispersion,
\begin{equation}
\Delta R_{xx}\propto -\frac{j_x}{k_\mathrm{F}}  \int d\Delta{k}\int d\alpha\int d\theta \left[ V_\mathrm{mag}(\theta,\alpha,\Delta{k})\left(\frac{\partial ^2 f}{\partial {k_x}^2} \right) v_\mathrm{F}\cos \alpha \right].
\label{D_Rxx}
	\stepcounter{equationnum}
	\tag{S\theequationnum}
\end{equation}
In this expression, we can exclude the contribution from $\tau^{(0)}$ and hence discuss the temperature dependence and $V_\mathrm{G}$ dependence of UMR of magnon origin. The calculation results in Fig. 3(d) in the main text are obtained by the numerical calculation of equation (\ref{D_Rxx}) assuming the simple circle Fermi surface of the Dirac cone.

\stepcounter{secnum}
\section*{Supplemental Material S\thesecnum\ \textbar\ Origin of deviation between the numerical calculations and the experimental results in Fig. 3}
Although the experimental results and numerical calculations give qualitative consistency, there is still quantitative deviation between them. The followings are the main causes of the deviation.
\begin{enumerate}
\item Relaxation time approximation \\
To calculate the scattering term at the right-hand side of Boltzmann equation (\ref{boltzman}) in the Supplemental Material S5, we used the relaxation time approximation so that
\begin{equation}
\left( \frac{dg(\bm{k})}{dt}\right)_\mathrm{coll}=-\frac{g(\bm{k})-g^0(\bm{k})}{\tau(\bm{k})}.
\label{relaxation_time}
	\stepcounter{equationnum}
	\tag{S\theequationnum}
\end{equation}
For a more serious calculation, it should be replaced by
\begin{equation}
\left( \frac{dg(\bm{k})}{dt}\right)_\mathrm{coll}=-\int \frac{d\bm{k'}}{(2\pi)^3}[W(\bm{k}|\bm{k'})g(\bm{k})(1-g(\bm{k'}))-W(\bm{k'}|\bm{k})g(\bm{k'})(1-g(\bm{k}))],
\label{wo_relaxation_time}
	\stepcounter{equationnum}
	\tag{S\theequationnum}
\end{equation}
where $W(\bm{k'}|\bm{k})$ is the scattering probability from $\bm{k}$ to $\bm{k'}$ \cite{ashcroft_mermin}. Since it is very difficult to carry out the calculation with equation (\ref{wo_relaxation_time}), we use relaxation time approximation in numerical calculation.

\item Spin wave approximation \\
Spin wave approximation is used in Supplemental Material S7 and S8 \cite{holstein_primakoff}. As mentioned in Supplemental Material S8, magnetic excitation is well-defined only at sufficiently low temperature than $T_\mathrm{C}$, therefore the theoretical calculation shows deviation at high temperatures.

\item Simplified band dispersion \\
For the sake of simplicity, we assumed a linear Dirac band dispersion. In fact, $k^2$ term and hexagonal warping ($k^3$ term) actually exist in the band dispersion in addition to the $k$-linear term \cite{TI_SS}. Although the linear Dirac band captures the essence of UMR, this also results in the deviation in the theoretical calculation.

\item Estimation of the spin wave gap \\
We put Zeeman term ($g\mu_\mathrm{B}B$) as a gap energy of magnon in the calculation. Since magnons with finite wavenumber is the source of scattering, we should take into account the band dispersion of magnon. To derive the band dispersion of magnon, however, it is necessary to estimate of the magnitude of the exchange interaction $J$, as well as to take into account the dipole-dipole interaction, which makes the calculation more complicated. Hence, as a first step, we considered only the Zeeman term this time.
\end{enumerate}

There is deviation between in the experimental results and the theoretical calculations because of some simplifications and approximations as described above. The essence of UMR is, however, captured by the theoretical formula; the change of the energy scale (peak position of $\Delta R_{xx}$) as a function of the magnetic field is well-reproduced in Fig. 3 in the main text. Improvement of the theoretical calculation without those simplifications and approximations remains as a future issue.

\end{document}